\documentclass[aps,prl,twocolumn,superscriptaddress]{revtex4}
\usepackage{graphicx}
\usepackage{ulem}
\usepackage{color}

\newcommand{\omegaDIFF}{\Delta \omega}

\newcommand{\ELO}{E_\mathrm{LO}}
\newcommand{\EI}{E_\mathrm{I}}
\newcommand{\EO}{E}

\newcommand{\envLO}{{\cal E}_\mathrm{LO}}
\newcommand{\envO}{{\cal E}}

\begin{document}


\title{Holographic laser Doppler ophthalmoscopy.}

\author{M. Simonutti}
\author{M. Paques}
\author{J. A. Sahel}

\affiliation{Institut de la Vision, Institut National de la Sant\'e et de la Recherche M\'edicale (INSERM), UMR-S 968.\\ Centre National de la Recherche Scientifique (CNRS) UMR 7210.\\ Universit\'e Pierre et Marie Curie (UPMC) 75012 Paris, France}

\author{M. Gross}
\affiliation{Laboratoire Kastler-Brossel de l'\'Ecole Normale
Sup\'erieure, CNRS UMR 8552, UPMC, 24 rue Lhomond 75005 Paris, France}

\author{B. Samson}
\author{C. Magnain}
\author{M. Atlan}
\affiliation{Institut Langevin, CNRS UMR 7587, INSERM U 979, Fondation Pierre-Gilles de Gennes, UPMC, Universit\'e Paris 7, ESPCI ParisTech, 10 rue Vauquelin, 75005 Paris, France.}


\date{\today}

\begin{abstract}

We report laser Doppler ophthalmoscopic fundus imaging in the rat eye, with near infrared heterodyne holography. Sequential sampling of the beat of the reflected radiation against a frequency-shifted optical local oscillator is made onto an array detector. Wide-field maps of fluctuation spectra in the 10 Hz to 25 kHz band exhibit angiographic contrasts in the retinal vascular tree without requirement of exogenous marker.

OCIS : 170.4470; 170.0110; 170.3340

\end{abstract}


\maketitle

Retinal blood flow plays a pivotal role in several blinding diseases such as diabetic retinopathy and vascular occlusions. While currently available optical instrumentation is well adapted for imaging retinal vessels, there
are still technological limitations of currently available methods for retinal blood flow measurement that impair their clinical applications. Self-beating intensity fluctuations from a laser spot focused in the eye fundus was demonstrated to enable retinal flow assessment \cite{Riva1979, Aizu1992, Fujii1994} non-invasively. It paved the way for the development of spatial scanning techniques such as confocal spot \cite{Michelson1996} or line \cite{Ferguson2004} scanning Doppler ophthalmoscopy. Yet, laser Doppler methods for retinal blood flow mapping are still limited in term of velocity resolution and mapping capabilities.

Optical coherence tomography, commonly used for structural retinal layers imaging, was translated to a depth-resolved functional Doppler-contrast technique \cite{ChenMilner1997b, Izatt1997}. Optical micro-angiography further improved depth-resolved Doppler measurements for eye fundus imaging \cite{AnWang2008}. Such techniques can enable quantitative and directional flow assessment, but vessel segmentation is required \cite{WangBower2007, MichaelyBachmann2007}.



Our approach to functional Doppler imaging is based on the detection of light fluctuations with an array detector. Multiply-scattered light yields Doppler spectra from which the directional information of flow is lost because of wave vector randomization. Nevertheless, it provides a spatially-resolved hemodynamic contrast in low-light illumination conditions. In this letter, we demonstrate experimentally the feasibility of wide-field holographic laser Doppler ophthalmoscopy, in vivo.

\begin{figure}[]
\centering
\includegraphics[width = 7.0 cm]{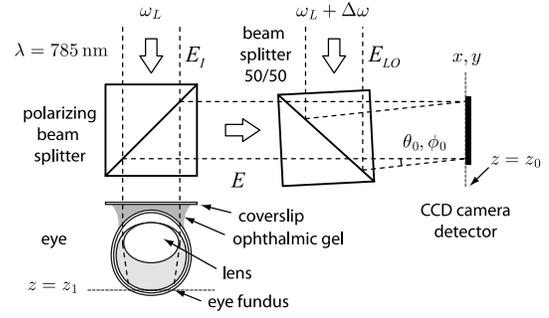}
\caption{Optical configuration. The eye fundus is illuminated in broad field through the dilated iris. The cornea curvature is compensated by a coverslip and ophthalmic aqueous gel.}\label{fig_setup}
\end{figure}

The experimental ophthalmoscope realized for this study is based on the heterodyne imaging scheme described in \cite{AtlanGrossVitalis2006}. It consists of a Mach-Zehnder laser interferometer in off-axis and frequency-shifting configuration. The detection scheme is sketched in Fig.\ref{fig_setup}. A laser diode provides the main near-infrared radiation at wavelength $\lambda = 785 \, \rm nm$, polarized linearly. In the object arm, a polarizing beam splitter cube is used to illuminate the preparation and collect the cross-polarized backscattered light component, in order to increase the relative weight of multiply scattered Doppler-shifted photons with respect to photons scattered once \cite{Schmitt1992}. Three adult rats were used for the preparations. Anesthesia was induced by intraperitoneal injection of 100 mg/kg ketamine and 25 mg/kg xylazine (both from Sigma-Aldrich). Topical tropicamide (CibaVision) was administered for pupil dilation. Each rat was placed on its side under the illumination beam. The head was supported so that the iris was perpendicular to the illumination axis. After administration of topical oxybuprocaine (CibaVision), a coverslip was applied on a ring surrounding the globe. Methylcellulose (Goniosol) was applied as contact medium. The incident optical field $\EI$ is expanded to form a plane wave. Illumination power within $5 \times 5 \, {\rm mm}^2$ at the object plane is $\sim$ 1 mW. In the reference arm (local oscillator : LO), an attenuator, a half wave plate, and a beam expander (not shown) are used to control the beam power, polarization angle, and to ensure a roughly flat LO illumination of the detector. The optical frequency detuning $\omegaDIFF$ between the two optical channels is imposed by acousto-optic modulators. The backscattered field $\EO$ is combined with the LO field $\ELO$ with a non-polarizing beam splitter cube. The detuning $\omegaDIFF$ shifts a component of interest of the scattered field temporal fluctuation spectrum within the actual camera bandwidth (temporal heterodyning). Moreover, a small angular tilt $\theta_0, \phi_0$ of $\sim 1 ^\circ$ ensures off-axis mixing conditions that shift the spatial frequency spectrum of the recorded object field (spatial heterodyning). The interference pattern $I= \left| \EO + \ELO \right|^2$ is measured by a Sony ICX 285AL charge-coupled device (CCD) array sensor (gain: 3.8 e/count), from which the central $1024 \times 1024$ pixels region is readout at $10 \, \rm Hz$. The detector is set $\sim 30 \, \rm cm$ away from the object plane. The recorded intensity pattern $I_n$ at time $t_n$ in the detector plane ($z=z_0$) is :
\begin{eqnarray}\label{eq_I_CMOS}
\nonumber I_n &=& \left| \envO \right|^2 + \left| \envLO \right|^2\\
\nonumber &+& \envO \envLO ^* {\rm e}^{- i (\omegaDIFF t_n +  \Delta k_{x} x + \Delta k_{y} y )} \\
&+& \envO ^* \envLO  {\rm e}^{+ i (\omegaDIFF t_n +  \Delta k_{x} x + \Delta k_{y} y )}
\end{eqnarray}
where $\envO$ and $\envLO$ are the complex envelopes of the scattered and LO field, respectively. $\Delta k_{x} = 2 \pi {\sin} ( \theta_0 ) / \lambda$, $\Delta k_{y} = 2 \pi {\sin} ( \phi_0 ) / \lambda$ are the projections of the difference between the LO and signal wave vectors in the transverse ($x,y$) plane. $^*$ denotes the complex conjugate. The two first terms of the right member of eq. \ref{eq_I_CMOS} are the self-beating (homodyne) contributions of $\EO$ and $\ELO$. The heterodyne signal of interest lies is the third term. The fourth term is the twin-image (ghost) contribution.

\begin{figure}[]
\centering
\includegraphics[width = 8.3 cm]{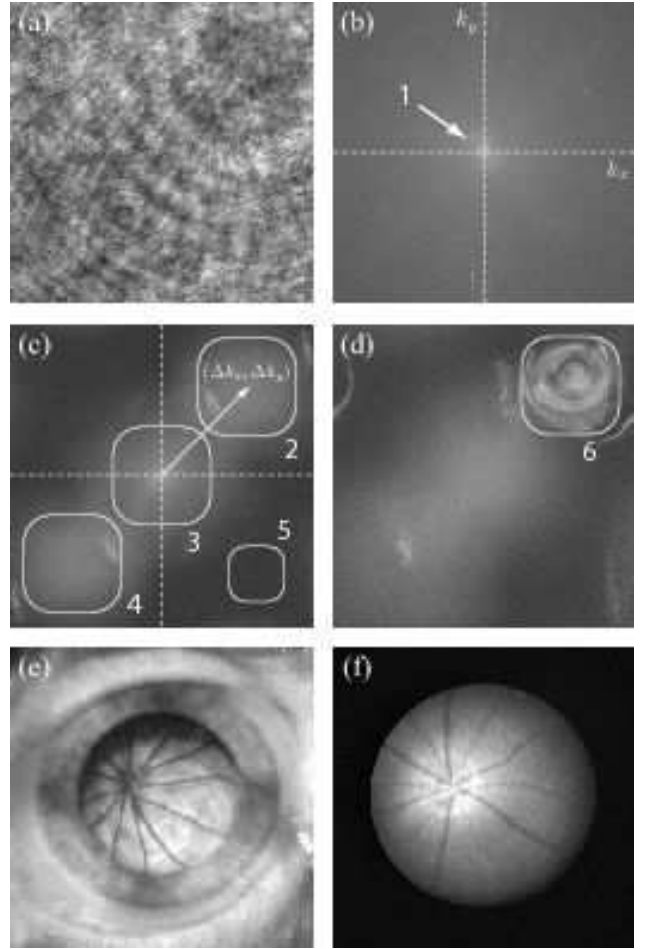}
\caption{Recorded interference pattern; $\Delta \omega / (2 \pi) = 10 \,\rm Hz$ (a). FFT of the recorded frame (b). The dominating noise (1) is gathered in the low spatial frequencies. FFT of the difference of two consecutive frames (c). Heterodyne contributions (2,4), homodyne contributions (3), noise (5). Image focus (6) by Fresnel transform (d). Magnified view of the retina (e). White-light endoscopic view of the retina (f). }\label{fig_080421_holographic_processing}
\end{figure}

\begin{figure}[]
\centering
\includegraphics[width = 8.3 cm]{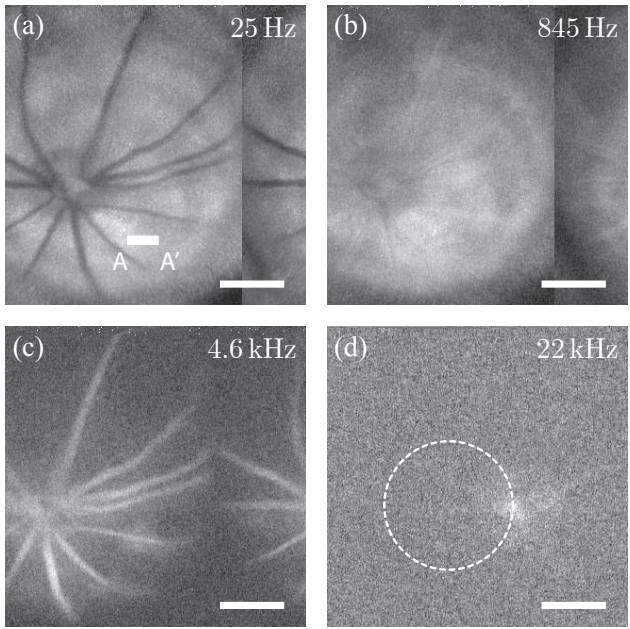}
\caption{Doppler fundus images at four frequency detunings $\omegaDIFF/(2 \pi)$. Optical power is displayed in logarithmic scale (white is for high signal). $[AA']$ indicates the measurement region of the lines reported in fig. \ref{fig_080428_spectres}(a). Scale bar, 1 mm.}\label{fig_4img_080428}
\end{figure}

\begin{figure}[]
\centering
\includegraphics[width = 8.3 cm]{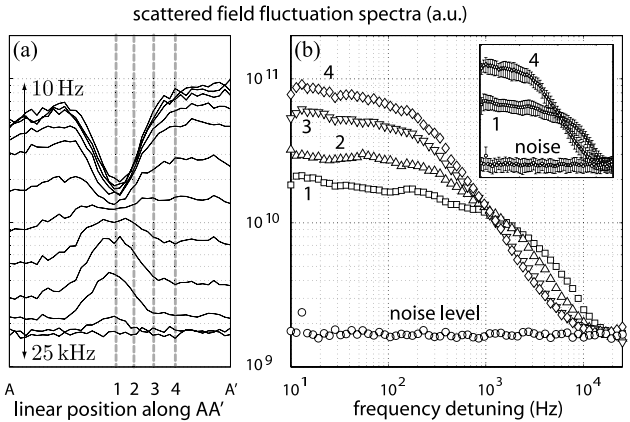}
\caption{(a) $\left< S \right>$ against position along $[AA']$ of fig. \ref{fig_4img_080428}(a), averaged over 15 pixels in the orthogonal direction. Traces correspond to logarithmically-spaced detuning frequencies from 10 Hz to 25 kHz. (b) $\left< S \right>$ against $\omegaDIFF$ at positions 1, 2, 3, 4 from the center of the vessel to its periphery.}\label{fig_080428_spectres}
\end{figure}

A typical interference pattern $I_n$ of $\EO$ beating against $\ELO$ is shown in fig. \ref{fig_080421_holographic_processing}(a). The magnitude of the fast spatial Fourier transform (FFT) of one recorded frame  $| {\rm FFT} (I_n) |^2 $ mainly carries the LO contribution in high heterodyne gain regime, when the optical power in the LO channel is larger than in the object channel [fig. \ref{fig_080421_holographic_processing}(b)]. Since the LO beam exhibits flat-field fluctuations, its self-beating contribution is gathered in the low frequency region (1) in reciprocal space $(k_x, k_y)$. Making the difference of two consecutive frames before (or after) applying the spatial Fourier transform yields $| {\rm FFT} (I_{n+1} - I_n) |^2$; it decreases substantially the relative weight of the LO self-beating term, showing up the object-against-LO beating term and the ghost term [regions (2) and (4), respectively, in fig. \ref{fig_080421_holographic_processing}(c)]. The object beam self-beating term also appears, in region 3. Detection noise is assessed in region (5). Because of off-axis geometry, the heterodyne signal contribution is shifted-away, by $\pm(\Delta k_{x},\Delta k_{y})$, from self-beating fields contributions. In on-axis geometry $(\Delta k_{x} = 0,\Delta k_{y} = 0)$, all interferometric terms would overlap spatially. The object field can hence be assessed with much better sensitivity than if measured in either on-axis-only or off-axis-only conditions and used for numerical reconstruction of the signal $S$ with a discrete Fresnel transform $S = | {\rm FFT} ( (I_{n+1} - I_n) {\rm e}^{i \frac{\pi}{\lambda \Delta z} (x^2 + y^2)} )|^2$, where the distance parameter used for free-space back-propagation of the optical field is $\Delta z = z_1 - z_0 = 26.6$ cm. Once accurate focus is found, the fundus image appears as reported in fig. \ref{fig_080421_holographic_processing}(d), in region 6, magnified in fig. \ref{fig_080421_holographic_processing}(e). These Doppler images are time-averaged $\left< S \right>$; angular brackets $\left< \, \right>$ account for averaging over 32 frames. The star-shaped vascular tree of the rat eye fundus is clearly visible and consistent with the white-light endoscopic image performed afterwards in the same preparation, displayed in fig. \ref{fig_080421_holographic_processing}(f).

Detuning $\omegaDIFF$ slows down selectively the drift rate of the set of fringes associated to a given Doppler component, and sets it within the actual camera bandwidth. Frequency-selective eye fundus images $\left< S \right>$ of a healthy rat at four LO detunings (25 Hz, 845 Hz, 4.6 kHz, and 22 kHz) are reported in Fig. \ref{fig_4img_080428}. A contrast reversal is observed between vessels and surrounding retinal tissue (and most likely in the choroid) from low to high frequency detunings [figs. \ref{fig_4img_080428}(a) and (c)]. Fluctuation spectra lines $\left< S \right>$ throughout a vessel cross section are reported in Fig. \ref{fig_080428_spectres}, as a function of position (a), and as a function of $\omegaDIFF$, at four locations within the vessel (b); typical spatial standard deviation of $\left< S \right>$ are shown in the inset. It can be estimated that choroidal flow significantly contributed to the signal along the retinal vessel section $[AA']$. Within this vessel, the Doppler spectrum is clearly broader from those of immediate surrounding areas, showing that the retinal flow predominantly contributed to the signal in this specific zone. The broadest Doppler signal, which is still visible at 22 kHz in fig. \ref{fig_4img_080428}(d) is found in the optic nerve head region (circled); it is interpreted as a consequence of vessels orientation and increased density in this area.


In conclusion, we have demonstrated the feasibility of holographic laser Doppler ophthalmoscopy with near infrared radiation.  The illumination power over the whole eye fundus can be kept low ($\sim$ 1 mW). High detection sensitivity, in moderate to high heterodyne gain regime, is achieved by spatiotemporal heterodyning, which enables an efficient rejection of parasitic interferometric contributions. Fluctuation spectra discrepancies below 25 kHz between the superficially vascularized regions and the surrounding retinal tissue layers provide an optical contrast suitable for angiographic mapping. A high reproducibility of the signal acquired over up to several hours, over repeated trials, and between different animals is observed. Potential applications are essentially the investigation of retinal and possibly choroidal vascular diseases. At the current stage there are yet limitations due to suboptimal lateral, depth, and time resolution, which for instance do not allow heart-beat related flow variations detection. Technical improvements are expected to circumvent these problems.

We acknowledge financial support from Agence Nationale de la Recherche (ANR-09-JCJC-0113 grant), Fondation Pierre-Gilles de Gennes (FPGG014 grant), Fondation Voir \& Entendre, R\'egion Ile-de-France, ESPCI and CNRS.


\bibliographystyle{unsrt}

\end{document}